\documentclass[10pt,a4paper]{article}
\usepackage{graphicx}
\newcommand{\Ap}{{A^\prime}}
\newcommand{\App}{{A^{\prime\prime}}}
\newcommand{\Bp}{{B^\prime}}
\newcommand{\Bpp}{{B^{\prime\prime}}}
\newcommand{\Cp}{{C^\prime}}
\newcommand{\Cpp}{{C^{\prime\prime}}}
\newcommand{\Pp}{{P^\prime}}
\date{}
\begin{document}

\date{}

\title{Global SSS space-time models: $M_a$ and $Q$\\
{\normalsize }}
\author{Ll. Bel\thanks{e-mail:  wtpbedil@lg.ehu.es}}

\maketitle

\begin{abstract}

To make sense of a global space-time model and to give a meaning to the coordinates that we use, a choice of a constant curvature space-metric of reference it is as much necessary as it is a choice of units of mass, length and time. The choice we make leads to contradict the belief that the exterior domain of a Static Spherically Symmetric (SSS) space-time model of finite radius $R$ depends only on the active mass $M_a$ of the source. In fact it depends on two parameters $M_a$ and a new one $Q$. We prove that both can be calculated as volume integrals extended over the whole space.

We integrate Einstein's equations numerically in two simple cases: assuming either that the source of perfect fluid has constant proper density or that the pressure depends linearly on the proper density. We confirm a preceding paper showing that very compact objects can have active  masses $M_a$ much greater than their proper masses $M_p$, and we conjecture that the mass point Fock's model can be understood as the limit of a sequence of compact models when both Q and its radius shrink to zero and the pressure equals the density.
.

\end{abstract}

\section{Global SSS space-time models}

Using Weyl's like decomposition, and obvious notations, we consider the line-element\,\footnote{$c=1,\  G=1$ are used throughout}:

\begin{equation}
\label{ds2}
ds^2=-A^2dt^2+A^{-2}d\bar{s}^2, \quad A=A(r)
\end{equation}
with:

\begin{equation}
\label{dsb2}
d{\bar s}^2=B^2dr^2+BCr^2d\Omega^2, \quad B=B(r),\ C=C(r)
\end{equation}

{\it Reference space-model}. We shall refer to this 3-dimensional metric as the space-model, and to restrict it as well as the coordinates used we shall require the condition:

\begin{equation}
\label{Cp}
\Cp=\frac{2}{r}(B-C)
\end{equation}
Introducing the Euclidean reference metric:

\begin{equation}
\label{dst2}
d{\tilde s}^2=dr^2+r^2d\Omega^2,
\end{equation}
we can write the equation above as follows:

\begin{equation}
\label{Gammag}
(\bar\Gamma^i_{jk}-\tilde\Gamma^i_{jk})g^{jk}=0, \quad i,j,k=1,2,3
\end{equation}
where $\bar\Gamma$ and $\tilde\Gamma$ are the connection symbols of (\ref{dsb2}) and (\ref{dst2}) in which case this equation becomes a restriction on the model but remains true whatever space coordinates we use. In particular Cartesian coordinates of (\ref{dst2}) become harmonic coordinates of (\ref{dsb2}) and this implies, as it is well known, that they are also harmonic coordinates of (\ref{ds2}).

{\it Einstein's tensor}.- We write below the non identically zero components of the Einstein's tensor:

\begin{eqnarray}
\label{S00}
S^0_0=\frac{2A\App}{B^2}-\frac{A^2\Bpp}{B^3}-\frac{3\Ap^2}{B^2}+\frac{4A\Ap}{rBC}-\frac{A^2}{r^2BC}+\frac{A^2}{r^2C^2}
-\frac{3A^2\Bp}{rB^2C}+\frac{5A^2\Bp^2}{4B^4}\\
\label{S11}
S^1_1=-\frac{A^2\Bp}{rB^2C}+\frac{A^2}{r^2BC}+\frac{\Ap^2}{B^2}-\frac{A^2}{r^2C^2}-\frac{A^2\Bp^2}{4B^4}\\
\label{S22}
S^2_2=S^3_3=-\frac{A^2\Bpp}{2B^3}+\frac{3A^2\Bp^2}{4B^4}-\frac{\Ap^2}{B^2}-\frac{A^2}{r^2BC}-\frac{A^2\Bp}{rB^2C}
+\frac{A^2}{r^2C^2}
\end{eqnarray}
where a prime means a derivative with respect to r. They have been somewhat simplified eliminating the derivatives of $C$ using (\ref{Cp}) and its derivative:

\begin{equation}
\label{Cpp}
\quad \Cpp=\frac{2}{r^2}(3C-3B+r\Bp)
\end{equation}

{\it The global model}.- the global perfect fluid models that we consider will  be solutions of Einstein's equations:

\begin{equation}
\label{Sij}
a:S^0_0=8\pi \rho, \quad b:S^1_1=-8\pi P, \quad c:S^2_2=S^3_3=-8\pi P
\end{equation}

A more convenient system of differential equations equivalent to the above one is the following:

\begin{eqnarray}
\label{Ode1}
\frac{2A\App}{B^2}-\frac{2\Ap^2}{B^2}+\frac{4A\Ap}{BCr}=8\pi (\rho+3P) \\
\label{Ode2}
-\frac12\frac{A^2\Bpp}{B^3}-\frac{\Ap^2}{B^2}+\frac34\frac{A^2\Bp^2}{B^4}-\frac{A^2}{BCr^2}+\frac{A^2}{C^2r^2}
-\frac{A^2\Bp}{B^2Cr}=-8\pi P \\
\label{Ode3}
\Pp=-\frac{\Ap}{A}(\rho+P) \\
\label{Ode4}
\Cp=\frac{2}{r}(B-C)
\end{eqnarray}

Eq. (\ref{Ode1}) is a linear combination of Eqs. $a$ and $c$. Eq. (\ref{Ode2}) is Eq. $c$ left unchanged. Eq. (\ref{Ode3}) is known to be equivalent to Eq. $b$ as a consequence of the conservation equations satisfied by the Einstein's tensor provided that $S^1_1=-8\pi P$ for $r=0$ . Eq. (\ref{Ode4}) is the same as Eq. (\ref{Cp})

As we shall see it is important to realize that Eq. (\ref{Ode1}) can be written as:

\begin{equation}
\label{LapU1}
\bar \Delta U=4\pi (\rho+3P)A^{-2}, \quad U=\ln(A)
\end{equation}
where $\bar \Delta$ is the Laplace operator of the space-model (\ref{dsb2}):

\begin{equation}
\label{LapU2}
\bar \Delta U =\frac{1}{\sqrt{\bar g}}\partial_i(\sqrt{\bar g}\bar g^{ij}\partial_jU), \quad
\bar g=\det|\bar g_{ij}|
\end{equation}
or explicitly:

\begin{equation}
\label{LapU3}
\bar \Delta U =\frac{1}{B^2Cr^2}\frac{d }{dr}\left(Cr^2\frac{dU}{dr}\right)
\end{equation}

\section{Boundary conditions}

To define a particular class of SSS models demands to be specific about the boundary conditions. This requires i) To state regularity conditions at the origin $r=0$. ii) To implement Lichnerowicz's continuity conditions at the boundary of the source, $r=R$, that we assume to be compact. And iii) to guarantee that from $r=R$ onwards $\rho$  and $P$ are zero and the solution has the correct asymptotic behavior.

{\it Regularity at the origin}.- To have continuity at the origin requires to have:

\begin{equation}
\label{Regp0}
\Ap_0=0, \quad   \Bp_0=0, \quad \Cp_0=0.
\end{equation}
and taking into account (\ref{Cp})there follows that we have to have also:
\begin{equation}
\label{RegB0}
B_0=C_0
\end{equation}
This means in particular that in a neighborhood of the origin we can write:

\begin{equation}
\label{A=A0+...}
A=A_0+\frac12 A_2r^2+O(r^3), \ B=B_0+\frac12 B_2r^2+O(r^3), \ C=B_0+\frac12 C_2r^2+O(r^3)
\end{equation}
with $C_2=1/2B_2$ to satisfy Eq. (\ref{Ode4}). It follows then from  (\ref{S11}) and (\ref{S22}) that at the origin we always have:

\begin{equation}
\label{S110}
(S^1_1)_0=(S_2^2)_0=(S_3^3)_0=-\frac{5}{4}\frac{A_0^2B_2}{B_0^3}
\end{equation}
that guarantees that the solutions of (\ref{Ode1})-(\ref{Ode4}) will satisfy Eq. $b$ of (\ref{Sij}).

{\it Matching conditions at $r=R$}.- The models  will be matched in the sense of Lichnerowicz, requiring the continuity of $A$, $B$ and $C$  and its first derivatives at $r=R$, $R$ being the radius of a compact source.

{\it Asymptotic conditions}: they will be the general asymptotic conditions of Schwarzschild's model compatible with (\ref{Cp}). They were derived in \cite{Bel}, and shall be used here\, \footnote{$Q$ in this paper is $QM_a^2$ in \cite{Bel}}:

\begin{eqnarray}
\label{Ainf}
A=1-\frac{M_a}{r}+\frac12\frac{M_a^2}{r^2}-\frac12\frac{M_a^3}{r^3}+O(1/r^4)\\
\label{Binf}
B=1-\frac23\frac{Q}{r^3}+O(1/r^4)\\
\label{Cinf}
C=1-\frac{M_a^2}{r^2}+\frac43\frac{Q}{r^3}+O(1/r^4)
\end{eqnarray}
where:

\begin{equation}
\label{MaLim}
M_a=\hbox{limit}_{r\rightarrow\infty}\Ap r^2
\end{equation}
is by definition the active mass of the source, and:

\begin{equation}
\label{QLim}
Q=\hbox{limit}_{r\rightarrow\infty}\frac12\Bp r^4
\end{equation}
is a new parameter that has already been considered in other contexts
\cite{Liu}-\cite{Bel}

\section{Integral expressions of $M_a$ and $Q$}

We define the proper mass of the source by:

\begin{equation}
\label{Mp}
M_p=4\pi \int_0^\infty \rho r^2\ dr
\end{equation}

Multiplying both members of (\ref{LapU1}) by $B^2Cr^2$ and integrating from $0$ to $\infty$ we get:

\begin{equation}
\label{Int}
\left(Cr^2\frac{dU}{dr}\right)_{\infty}-\left(Cr^2\frac{dU}{dr}\right)_0=4\pi \int_0^\infty (\rho+3P)A^{-2}BCr^2\ dr
\end{equation}
that using (\ref{Ainf}) and (\ref{Cinf}) becomes the well-known Tolman's formula \cite{Tolman}:

\begin{equation}
\label{Tolman}
M_a=4\pi \int_0^R(\rho+3P)A^{-2}BCr^2\ dr
\end{equation}

On the other hand, integrating by parts the first two terms of $S^0_0r^2$ over all space, and taking into account the conditions stated above we obtain:

\begin{equation}
\label{Ma=Mp+...}
M_a=M_p+4\pi \int_0^\infty\sigma_\rho r^2\ dr
\end{equation}
where:

\begin{equation}
\label{sigmarho}
\sigma_\rho=\frac{5\Ap^2}{B^2}-\frac{6A\Ap\Bp}{B^3}+\frac{4A\Ap}{rB^2}+\frac{7A^2\Bp^2}{4B^4}
-\frac{2A^2\Bp}{rB^3}-\frac{4A\Ap}{rBC}+\frac{A^2}{BCr^2}-\frac{A^2}{C^2r^2}+\frac{3A^2\Bp}{B^2Cr}
\end{equation}

Similarly integrating by parts the first term of $(S^2_2-S^1_1)r^4$ and using the boundary conditions above we obtain:

\begin{equation}
\label{Q}
Q=\int_0^\infty \sigma_p r^2\ dr
\end{equation}
where:

\begin{equation}
\label{sigmaP}
\sigma_p=-\frac{2r^2\Ap^2}{B^2}-\frac{2A^2}{BC}-\frac{2r^2A^2\Bp^2}{B^4}
+\frac{2A^2}{C^2}+\frac{r^2A\Ap\Bp}{B^3}+\frac{2rA^2\Bp}{B^3}
\end{equation}

\section{Numerical models: Examples}

We consider two cases based on the following relationships between the density $\rho$ and the pressure $P$.
\begin{itemize}
\item Case I: we assume that the density $\rho$ is constant and positive from $r=0$ to $r=R$, where $P=0$;
\item Case II: we assume that $P=k\rho- \eta$ from $r=0$ to $r=R$, where P is a small fraction of an assumed $P_0$ value at the origin. $k=1/3$ or $k=1$  and $\eta$ is a positive constant included to guarantee that the pressure reaches the value zero at some radius $R$;
\end{itemize}

The initial initial conditions will be:

\begin{equation}
\label{Ap0}
\Ap_0=\Bp_0=\Cp_0=0 \quad B_0=C_0=1, \quad \rho0=(8\pi)^{-1}
\end{equation}
compatible with the regularity conditions discussed above.

The following coordinate transformations:

\begin{equation}
\label{mu}
t=\mu\bar t, \quad r=\nu\bar r
\end{equation}
modifies $A$, $B$ and $C$ as follows:

\begin{equation}
\label{Abar}
\bar A=\mu A, \quad \bar B=\nu\mu B, \quad \bar C=\nu\mu C,
\end{equation}
but leaves invariant (\ref{Cp}) as well as  (\ref{S00})-(\ref{S22}) and therefore $\rho$ and $P$. A covariant property that will be used below.

The integration proceeds in two steps and two runs. First of all we integrate Eqs. (\ref{Ode1})-(\ref{Ode4}) until $P$ reaches a chosen estimated small value at some $r=R$. Beyond that we set $\rho=0$, $P=0$ and the integration proceeds until the quantity $\epsilon=rAp$ reaches a chosen estimated small value.
Let $A_\epsilon$ and Let $B_\epsilon$ be the corresponding values of $A$ and $B$.

The second round consists in integrating in two steps the same differential Eqs. (\ref{Ode1})-(\ref{Ode4})  with the same initial values of $\rho_0$ and $P_0$ but with the following different initial conditions:

\begin{equation}
\label{Abar0}
\bar A_0=\frac{A_0}{A_\epsilon}, \quad \bar B_0=\frac{A_0}{B_\epsilon}
\end{equation}
which is one of the coordinate transformations considered in (\ref{mu}), with:

\begin{equation}
\label{mu}
\mu=A_\epsilon^{-1}, \quad \lambda= A_\epsilon B_\epsilon^{-1}
\end{equation}

{\it Examples}.- The parameter $\eta$ in Case II is $0.01$ for $k=1/3$ and $0.10$ for $k=1$, and Eqs. (\ref{Ode1})-(\ref{Ode4}) have been integrated until $P$ became of the order of $1.0\,e-8$.

The following tables where $\lambda$ is the compactness parameter:

\begin{equation}
\label{lambda}
\lambda=\frac{M_a}{R}
\end{equation}
summarize some of the informative results:
\vspace{1cm}

Case I:
\vspace{1cm}

\begin{tabular}{|l|l|l|l|l|l|}
\hline
\\[-2ex]
$\quad k$ & $\quad R$ & $\quad M_p$ & $\quad M_a$ & $\quad \lambda $ & $\quad Q$
\\[.5ex]
\hline
1/3 & \ .92 & \ .13 & \ .36 & \ .39& \ .02 \\
\ 1 & \ .91 & \ .13 & \ .57 & \ .62& \ .43 \\
\hline
\end{tabular}
\vspace{1cm}

Case II:
\vspace{1cm}

\begin{tabular}{|l|l|l|l|l|l|l|}
\hline
\\[-2ex]
$\quad k$ & $\quad \eta$ & $\quad R$ & $\quad M_p$ & $\quad M_a$ & $\quad \lambda$ & $\quad Q$
\\[.5ex]
\hline
1/3 & \ 0.01 & \ 3.21 & \ 0.332 & \ 1.194 & \ .030 & \ 11.4 \\
\ 1 & \ 0.10 & \ 2.10 & \ 0.240 & \ 1.222 & \ .046 & \ 11.7 \\
\hline
\end{tabular}

\section*{Appendix}

This appendix is an update of our reference \cite{Bel23} that is meant to prove that Fock's model \cite{Fock} of the Schwarzschild's exterior solution can be considered as a legitimate limit model of regular SSS models when the proper volume that contains the proper mass shrinks to zero.

For this model we have:

\begin{equation}
\label{A.1}
A=\sqrt{\frac{r-M_a}{r+M_a}}, \quad B=1, \quad C= 1-\frac{M_a^2}{r^2}
\end{equation}
and evaluating (\ref{sigmarho})and (\ref{sigmaP}) we obtain:

\begin{equation}
\label{A.2}
\sigma_\rho=\frac{4M_a^2}{r(r+m)},\quad \sigma_p=0
\end{equation}
and therefore using (\ref{Ma=Mp+...}) and :

\begin{equation}
\label{A.3}
4\pi\int_{M_a}^\infty \sigma_\rho r^2\,dr=\frac{3}{4}M_a
\end{equation}
we get:

\begin{equation}
\label{A.4}
M_p=\frac14 M_a, \quad Q=0.
\end{equation}
The first of these formulas suggest that the source of Fock's model is:

\begin{equation}
\label{A.5}
\rho=\frac14 M_a\delta(r-M_a)
\end{equation}
where the Dirac density will be defined by the two properties:

\begin{equation}
\label{A.6}
r>Ma \Rightarrow \delta(r-M_a)=0,\quad \hbox{and}\quad 4\pi\int_{M_a}^\infty \delta(r-M_a)r^2\, dr=1
\end{equation}
Tolman's formula (\ref{Tolman}) for general regular models and Eq. (\ref{A.5}) above suggest that if the proper density for Fock's model contributes to only one third of the active mass then the pressure must be:

\begin{equation}
\label{A.7}
P=\frac14 M_a\delta(r-M_a)
\end{equation}
so that:

\begin{equation}
\label{A.8}
P=\rho \quad \hbox{and} \quad  \rho+3P=M_a\delta(r-M_a)
\end{equation}

Now, let us take a look to the geometry (\ref{dsb2}) corresponding to the coefficients $B$ and $C$ above. We have:

\begin{equation}
\label{A.9}
\sqrt{\det\bar g_r}=B^2C\sin\theta=(r^2-M_a^2)\sin\theta
\end{equation}
that becomes zero when $r=M_a$. This means that the system of coordinates whose meaning is that derived from the metric of reference (\ref{dst2}) is not to be trusted when discussing properties of the model in a neighborhood of $r=M_a$.

What we do below is more than we need to discuss this point but we prefer here to deal with a global condition instead of a local one as we did in \cite{Bel23} and introduce a system of coordinates:

\begin{equation}
\label{A.10}
y_1=u(r)\sin\theta\cos\phi,\ y_2=u(r)\sin\theta\sin\phi, \ y_3=u(r)\cos\theta
\end{equation}
$u$ being a function of r such that:

\begin{equation}
\label{A.11}
\sqrt{\det\bar g_y}=1 \Leftrightarrow \sqrt{\det\bar g_u}=u^2
\end{equation}
Using the polar form of the condition, a short calculation proves that the function $u$ will be given by:

\begin{equation}
\label{A.12}
u=(r^3-3M_a^2r+2M_a^3)^{1/3}
\end{equation}
so that:

\begin{equation}
\label{A.13}
u(M_a)=0
\end{equation}

The other way around, to get the inverse function $r(u)$ we need to solve the third degree algebraic equation:

\begin{equation}
\label{A.14 }
r^3-3M_a^2r+2M_a^3-u^3=0
\end{equation}
The discriminant of this equation is positive if $u>2^{2/3}M_a$ and negative otherwise down to $u=0$. In the first case there exist only one real solution, namely:

\begin{equation}
\label{A.15 }
r=\left(-M_a^3+\frac12 u^3+\frac12\sqrt{-4M_a^3u^3+u^6}\right)^{1/3}
+ \left(-M_a^3+\frac12 u^3-\frac12\sqrt{-4M_a^3u^3+u^6}\right)^{1/3}
\end{equation}
that has an asymptote $r=u$, while in the second case there are three real functions. But there is only one for which $r$ is a positive increasing function of u, namely:

\begin{equation}
\label{A.16 }
r=2M_a\sin\left(\frac16\pi+\frac13\arccos\left(1-\frac12\frac{u^3}{M_a^3}\right)\right)
\end{equation}
whose behavior near $u=0$ is:

\begin{equation}
\label{A.17 }
r=M_a+\frac13\frac{\sqrt{3}u^{3/2}}{\sqrt{M_a}}-\frac{1}{18}\frac{u^3}{M_a^2}+O(u^{9/2})
\end{equation}
The two solutions here considered joint smoothly at $u=2^{2/3}M_a$.

Using the radial coordinate $u$ instead of $r$ the space-model (\ref{dsb2}) becomes:

\begin{equation}
\label{A.18 }
d{\bar s}^2=B_u^2du^2+B_u^{-1}u^2d\Omega^2
\end{equation}
where:

\begin{equation}
\label{A.19 }
B_u=\frac{dr}{du}=\frac{\cos\left(\frac16\pi+\frac13\arccos\left(1-\frac12\frac{u^3}{M_a^3}\right)\right)u^2}
{M_a^2\sqrt{1-\left(1-\frac12\frac{u^3}{M_a^3}\right)^2}}
\end{equation}
and $B_u^2$ can be approximated in a neighborhood of u=0 by:

\begin{equation}
\label{A.20 }
B_u^2=\frac34\frac{u}{M_a}-\frac16\frac{\sqrt{3}u^{5/2}}{M_a^{5/2}}+O(u^4)
\end{equation}
while $U_u$ and $dU_u/du$ can be approximated by:

\begin{equation}
\label{A.21 }
U_u=\frac12 \ln\left(\frac16\frac{\sqrt{3}}{M_a^3/2}\right)+\ln(u))-\frac19\frac{\sqrt(3)u^{3/2}}{m^{3/2}}+O(u^3)
\end{equation}
and:

\begin{equation}
\label{A.22 }
\frac{dU_u}{du}=\frac34\frac{u}{M_a}-\frac16\frac{\sqrt{3}\sqrt{u}}{M_a^{3/2}}+O(u^2)
\end{equation}

Let us now consider again Eq. (\ref{LapU3}) that becomes now, for Fock's model, the Laplace equation:

\begin{equation}
\label{A.23 }
\bar\Delta U_u=\frac{1}{u^2}\frac{d }{du}\left(u^2B_u^{-2}\frac{dU_u}{du}\right)=0
\end{equation}
that holds for $u>0$ but it is actually undefined for $u=0$.

Let us define $U_u$ as the linear functional:

\begin{equation}
\label{A.24 }
<U_u,\varphi>=4\pi\,\hbox{limit}_{\epsilon\rightarrow 0}\int_\epsilon^\infty U_u\varphi u^2du
\end{equation}
where $\varphi$ is any function such that itself and all its derivatives become zero beyond some finite value of $u$.
As a distribution we shall have then:

\begin{equation}
\label{A.25 }
<\bar\Delta U_u,\varphi>=4\pi\,\hbox{limit}_{\epsilon\rightarrow 0}\int_\epsilon^\infty U_u\bar\Delta\varphi u^2\,du
\end{equation}
or:

\begin{equation}
\label{A.26}
<\bar\Delta U_u,\varphi>=4\pi\,\hbox{limit}_{\epsilon\rightarrow 0}\int_\epsilon^\infty U_u\frac{d }{du}\left(u^2B_u^{-2}\frac{d\varphi}{du}\right)\,du
\end{equation}
Using now this particular case of Green's formula:

\begin{equation}
\label{A.27 }
U_u\frac{d }{du}\left(u^2B_u^{-2}\frac{d\varphi}{du}\right)-
\varphi\frac{d }{du}\left(u^2B_u^{-2}\frac{dU_u}{du}\right)=
\frac{d }{du}\left(u^2B_u^{-2}U_u\frac{d\varphi}{du}\right)-
\frac{d }{du}\left(u^2B_u^{-2}\varphi\frac{dU_u}{du}\right)
\end{equation}
and taking into account (\ref{A.26}) we have:

\begin{equation}
\label{A.28 }
<\bar\Delta U_u,\varphi>=4\pi\,\hbox{limit}_{u\rightarrow 0}\left(u^2B_u^{-2}\left(U_u\frac{d\varphi}{du}-\varphi\frac{dU_u}{du}\right)\right)
\end{equation}
Now since:

\begin{equation}
\label{A.29 }
\hbox{limit}_{u \rightarrow 0}(u^2B_u^{-2}U_u)=0 \quad \hbox{and}
\quad \hbox{limit}_{u\rightarrow 0}\left(u^2B_u^{-2}\frac{dU_u}{du}\right)=M_a
\end{equation}
we conclude that:

\begin{equation}
\label{A.30 }
<\bar\Delta U_u,\varphi>=4\pi M_a\varphi(0), \ \hbox{or} \ \bar\Delta U_u=4\pi\frac12 M_a\delta(u)
\end{equation}





\end{document}